\documentclass[preprintnumbers,amsmath,amssymbm,prl]{revtex4}
\usepackage{epsfig}
\usepackage{graphicx}

\begin{document}
\title{On the instability regime of the rotating Kerr spacetime to massive scalar perturbations}
\author{Shahar Hod}
\affiliation{The Ruppin Academic Center, Emeq Hefer 40250, Israel}
\affiliation{ } \affiliation{The Hadassah Institute, Jerusalem
91010, Israel}
\date{\today}

\begin{abstract}
\ \ \ The instability of rotating Kerr black holes due to massive
scalar perturbations is investigated. It is well known that a
bosonic field impinging on a Kerr black hole can be amplified as it
scatters off the hole. This superradiant scattering occurs for
frequencies in the range $\omega<m\Omega$, where $\Omega$ is the
angular frequency of the black hole and $m$ is the azimuthal
harmonic index of the mode. If the incident field has a non-zero
rest mass, $\mu$, then the mass term effectively works as a mirror,
reflecting the scattered wave back towards the black hole. The wave
may bounce back and forth between the black hole and some turning
point amplifying itself each time. This may lead to a dynamical
instability of the system, a phenomena known as a ``black-hole
bomb". In this work we provide a bound on the instability regime of
rotating Kerr spacetimes. In particular, we show that Kerr black
holes are stable to massive perturbations in the regime
$\mu\geq\sqrt{2}m\Omega$.
\end{abstract}
\bigskip
\maketitle


If a black hole is perturbed in some small way, will the
perturbation die away over time? Or will it grow exponentially until
it can no longer be considered a perturbation and hence demonstrate
the instability of the black hole? The fundamental role played by
black holes in many areas of modern physics (astrophysics,
high-energy physics, condensed matter physics) makes it highly
important to study the nature of their stability.

Regge and Wheeler \cite{RegWheel} have provided evidence for the
stability of the spherically symmetric Schwarzschild black hole. In
particular, they have shown that if a Schwarzschild black hole is
perturbed, then the perturbation will oscillate and damp out over
time \cite{RegWheel,Vish}. This implies that perturbation fields
which propagate in the black-hole exterior would either be radiated
away to infinity or swallowed by the black hole.

It is well known that realistic stellar objects generally rotate
about their axis and are therefore not spherical. Thus, an
astrophysically realistic model of wave dynamics in black-hole
spacetimes must involve a non-spherical background geometry with
angular momentum. The stability question of rotating Kerr black
holes is more complicated than the spherically-symmetric
Schwarzschild case. Press and Teukolsky \cite{PressTeu1,Teu} have
shown that rotating black holes are stable under free gravitational
perturbations (see also \cite{Hodp1} and references therein).
However, the superradiance effect may change this conclusion.
Superradiant scattering is a well-known phenomena in quantum systems
\cite{Mano,Grein} as well as in classical ones \cite{Zel,Car}.
Considering a wave of the form $e^{im\phi}e^{-i\omega t}$ (where $m$
and $\omega$ are real) incident upon a rotating object whose angular
velocity is $\Omega$, one finds that if the frequency $\omega$ of
the incident wave satisfies the relation
\begin{equation}\label{Eq1}
\omega<m\Omega\  ,
\end{equation}
then the scattered wave is amplified.

A bosonic field impinging upon a rotating Kerr black hole can be
amplified if the superradiance condition (\ref{Eq1}) is satisfied,
where in this case
\begin{equation}\label{Eq2}
\Omega={a\over {r^2_++a^2}}
\end{equation}
is the angular velocity of the black-hole horizon. Here $r_+$ and
$a$ are the horizon radius and the angular momentum per unit mass of
the black hole, respectively. The energy radiated away to infinity
may actually exceed the energy present in the initial perturbation.
Feeding back the amplified scattered wave, one can gradually extract
the rotational energy of the black hole. Press and Teukolsky
suggested to use this mechanism to build a {\it black-hole bomb}
\cite{PressTeu2}: If one surrounds the black hole by a reflecting
mirror, the wave will bounce back and forth between the black hole
and the mirror amplifying itself each time. Thus, the total energy
extracted from the black hole will gradually grow.

Remarkably, nature sometimes provides its own mirror \cite{Car}: If
one considers a {\it massive} scalar field with mass ${\cal M}$
scattered off a rotating black hole, then for $\omega<\mu\equiv
{\cal M}G/\hbar c$ the mass term effectively works as a mirror
\cite{Dam,Zour,Det,Furu,Dolan,CarYo,HodHod}. The physical idea is to
consider a wave packet of the massive field in a bound orbit around
the black hole \cite{Zour,Det}. The gravitational force binds the
field and keeps it from escaping to infinity. At the event horizon
some of the field goes down the black hole, and if the frequency of
the wave is in the superradiance regime (\ref{Eq1}) then the field
is amplified. In this way the field is amplified at the horizon
while being bound away from infinity. This may lead to a dynamical
instability of the system
\cite{Dam,Zour,Det,Furu,Dolan,CarYo,HodHod}.

Former studies \cite{Dam,Zour,Det,Furu,Dolan,CarYo,HodHod} of the
dynamics of massive scalar fields in rotating Kerr spacetimes have
shown that the superradiant instability (the black hole bomb
mechanism) depends on two parameters:
\begin{itemize}
\item{The angular frequency $\Omega$ of the black hole.}
\item{The dimensionless product of the black hole mass $M$ and the field
mass $\mu$. The product $M\mu$ is actually the ratio of the black
hole size to the Compton wavelength associated with the rest mass of
the field. (We shall henceforth use natural units in which
$G=c=\hbar=1$. In these units $\mu$ has the dimensions of
$1/$length.)}
\end{itemize}
In particular, it has been proved \cite{Beyer1} that the solutions
of the massive scalar equation in the Kerr spacetime are stable in
the regime
\begin{equation}\label{Eq3}
\mu\geq m\Omega\sqrt{1+{{2M}\over{r_+}}+{{a^2}\over{r^2_+}}}\  .
\end{equation}
Recently, an improved bound for $\mu$ above which the solutions of
the massive scalar equation are stable was given in \cite{Beyer2}:
\begin{equation}\label{Eq4}
\mu\geq m\Omega\sqrt{1+{{2M}\over{r_+}}}\  .
\end{equation}
Below we shall provide a {\it stronger} bound for the stability
regime of rotating Kerr black holes: we shall show that the
solutions of the massive Klein-Gordon equation in the Kerr spacetime
are stable in the regime
\begin{equation}\label{Eq5}
\mu\geq m\Omega\cdot\sqrt{2}\
\end{equation}
for all $a$ values.

The physical system we consider consists of a massive scalar field
coupled to a rotating Kerr black hole. The dynamics of a scalar
field $\Psi$ of mass $\mu$ in the Kerr spacetime \cite{Kerr} is
governed by the Klein-Gordon equation
\begin{equation}\label{Eq6}
(\nabla^a \nabla_a -\mu^2)\Psi=0\  .
\end{equation}
One may decompose the field as
\begin{equation}\label{Eq7}
\Psi_{lm}(t,r,\theta,\phi)=e^{im\phi}S_{lm}(\theta;a\omega)R_{lm}(r;a\omega)e^{-i\omega
t}\ ,
\end{equation}
where $(t,r,\theta,\phi)$ are the Boyer-Lindquist coordinates
\cite{Kerr}, $\omega$ is the (conserved) frequency of the mode, $l$
is the spheroidal harmonic index, and $m$ is the azimuthal harmonic
index with $-l\leq m\leq l$. (We shall henceforth omit the indices
$l$ and $m$ for brevity.) With the decomposition (\ref{Eq7}), $R$
and $S$ obey radial and angular equations both of confluent Heun
type coupled by a separation constant $K(a\omega)$
\cite{Heun,Flam,Fiz1,Fiz2,Fiz3}. The sign of $\omega_I$ determines
whether the solution is decaying $(\omega_I<0)$ or growing
$(\omega_I>0)$ in time. Remembering that any instability must set in
via a real-frequency mode \cite{Hartle} we shall consider here modes
with $|\omega_I|\ll \omega_R$. (It should be emphasized that Ref.
\cite{Hartle} considered only the massless case.)

The angular functions $S(\theta;a\omega)$ are the spheroidal
harmonics which are solutions of the angular equation
\cite{Teuk,Flam,Abram,Hodcen,Stro}
\begin{equation}\label{Eq8}
{1\over {\sin\theta}}{\partial \over
{\partial\theta}}\Big(\sin\theta {{\partial
S}\over{\partial\theta}}\Big)+\Big[K-a^2(\omega^2-\mu^2)+a^2(\omega^2-\mu^2)\cos^2\theta-{{m^2}\over{\sin^2\theta}}\Big]S=0\
.
\end{equation}
The angular functions are required to be regular at the poles
$\theta=0$ and $\theta=\pi$. These boundary conditions pick out a
discrete set of eigenvalues $\{K_{lm}\}$ labeled by the integers $l$
and $-l\leq m\leq l$.

The radial Klein-Gordon equation is given by
\cite{Teuk,Hodcen,Stro,Carter}
\begin{equation}\label{Eq9}
\Delta{{d}\over{dr}}\Big(\Delta{{dR}\over{dr}}\Big)+UR=0\  ,
\end{equation}
where $\Delta\equiv r^2-2Mr+a^2$, and
\begin{equation}\label{Eq10}
U\equiv [(r^2+a^2)\omega-ma]^2+\Delta[2ma\omega-K-\mu^2(r^2+a^2)]\
.
\end{equation}
The zeroes of $\Delta$, $r_{\pm}=M\pm (M^2-a^2)^{1/2}$, are the
black hole (event and inner) horizons.

We are interested in solutions of the radial equation (\ref{Eq9})
with the physical boundary conditions of purely ingoing waves at the
black-hole horizon (as measured by a comoving observer) and a
decaying (bounded) solution at spatial infinity \cite{Zour,Dolan}.
That is,
\begin{equation}\label{Eq11}
R \sim e^{-i (\omega-m\Omega)y}\ \ \text{ as }\ r\rightarrow r_+\ \
(y\rightarrow -\infty)\ ,
\end{equation}
and
\begin{equation}\label{Eq12}
R \sim {1\over r}e^{-\sqrt{\mu^2-\omega^2}y}\ \ \text{ as }\
r\rightarrow\infty\ \ (y\rightarrow \infty)\  ,
\end{equation}
where the ``tortoise" radial coordinate $y$ is defined by
$dy=[(r^2+a^2)/\Delta]dr$ \cite{Notered}. For frequencies in the
superradiant regime (\ref{Eq1}), the boundary condition (\ref{Eq11})
describes an outgoing flux of energy and angular momentum from the
rotating black hole \cite{Zour,Dolan}. Note also that a bound state
(a state decaying exponentially at spatial infinity) is
characterized by
\begin{equation}\label{Eq13}
\omega^2<\mu^2\  .
\end{equation}
The boundary conditions (\ref{Eq11})-(\ref{Eq13}) single out a
discrete set of resonances $\{\omega_n\}$ which correspond to the
bound states of the massive field \cite{Zour,Dolan}. (We note that,
in addition to the bound states of the massive field, the field also
has an infinite set of discrete quasinormal resonances
\cite{Will,ZhKo,Hodmassive,Massfr} which are characterized by
outgoing waves at spatial infinity.)

It is convenient to define a new radial function $\psi$ by
\begin{equation}\label{Eq14}
\psi\equiv \Delta^{1/2}R\  ,
\end{equation}
in terms of which the radial equation (\ref{Eq9}) can be written in
the form of a Schr\"odinger-like wave equation
\begin{equation}\label{Eq15}
{{d^2\psi}\over{dr^2}}+(\omega^2-V)\psi=0\  ,
\end{equation}
where
\begin{equation}\label{Eq16}
\omega^2-V={{U+M^2-a^2}\over{\Delta^2}}\  .
\end{equation}
Below we shall need the asymptotic form of the effective potential
which is given by
\begin{equation}\label{Eq17}
V(r)=\mu^2-{{4M\omega^2-2M\mu^2}\over{r}}+O\Big({{1}\over{r^2}}\Big)\
.
\end{equation}

As discussed above, the instability of the Kerr spacetime to massive
scalar perturbations is a consequence of massive modes which are
trapped inside the effective potential well outside the black hole.
These modes are superradiantly amplified in the ergo-region near the
black-hole horizon. Two ingredients are therefore required in order
to trigger the instability:
\begin{itemize}
\item{The existence of a trapping potential well outside the black hole.}
\item{The existence of an ergo-region, in which superradiant amplification of the waves
takes place.}
\end{itemize}

For the effective potential to have a trapping well, its asymptotic
derivative must be positive ($V'\to 0^+$ as $r\to\infty$). Taking
cognizance of Eqs. (\ref{Eq13}) and (\ref{Eq17}), one finds that
bound states may exist for modes in the regime
\begin{equation}\label{Eq18}
{{\mu^2}\over{2}}<\omega^2<\mu^2\  .
\end{equation}
In Table \ref{Table1} we test the validity of the inequality
${{\mu^2}\over{2}}<\omega^2$ for the bound states of the massive
scalar field. For this conformation we use numerical results for the
bound state frequencies of massive scalar fields in Kerr spacetimes,
see also \cite{Dolan}. From Table \ref{Table1} one indeed learns
that all bound states (for all $a$ values) conform to the inequality
(\ref{Eq18}).

\begin{table}[htbp]
\centering
\begin{tabular}{|c|c|c|c|}
\hline
$a/M$ & $M\mu^{*}$ & $\text{min}_{M\mu}\{2\omega^2/\mu^2\}$ \\
\hline
\ 0.0\ \ & \ 0.943\ \ &\ 1.677\ \\
\ 0.3\ \ & \ 1.005\ \ &\ 1.618\ \\
\ 0.5\ \ & \ 1.032\ \ &\ 1.580\ \\
\ 0.7\ \ & \ 1.021\ \ &\ 1.522\ \\
\ 0.9\ \ & \ 0.933\ \ &\ 1.473\ \\
\ 0.99\ \ & \ 0.869\ \ &\ 1.466\ \\
\ 0.999\ \ & \ 0.867\ \ &\ 1.465\ \\
\hline
\end{tabular}
\caption{Massive scalar bound states of rotating Kerr black holes.
The data shown is for the mode $l=m=1$ and for various values of the
rotation parameter $a$, see also \cite{Dolan}. We display the ratio
$\text{min}_{M\mu}\{2\omega^2/\mu^2\}$ for the numerically computed
frequencies. (We also display the dimensionless product $M\mu^{*}$,
where the ratio $2\omega^2/\mu^2$ attains its minimum.) One finds
$2\omega^2/\mu^2>1$ for all $a$ values, in agreement with the
inequality (\ref{Eq18}).} \label{Table1}
\end{table}

In addition to the requirement (\ref{Eq18}) for the existence of a
trapping well, the superradiance condition for amplification of the
scattered waves is given by [see Eq. (\ref{Eq1})]
\begin{equation}\label{Eq19}
\omega<m\Omega\  .
\end{equation}
One therefore realizes that the two conditions for superradiant
instability [see Eqs. (\ref{Eq18}) and (\ref{Eq19})] can be
satisfied in the restricted regime $\mu/\sqrt{2}<\omega<m\Omega$. In
other words, the dynamics of the massive scalar field is expected to
be stable in the complementary regime
\begin{equation}\label{Eq20}
\mu\geq\sqrt{2}\cdot m\Omega\  .
\end{equation}
It is worth emphasizing that the new bound (\ref{Eq20}) coincides
with the former bound (\ref{Eq4}) in the slow rotation $a\to 0$
limit (in which case $r_+\to 2M$). However, for finite $a$ values
the bound (\ref{Eq20}) is {\it stronger} than the bound (\ref{Eq4})
\cite{Noteunbound}.

In summary, we have studied the instability of rotating Kerr black
holes to massive scalar perturbations. In particular, we have shown
that the two conditions which are necessary for the existence of the
superradiant instability [namely: (1) the existence of a trapping
well, and (2) superradiant amplification of the trapped modes] can
only be satisfied in the restricted regime $\mu<\sqrt{2}\cdot
m\Omega$. Thus, the dynamics of the massive scalar field in the Kerr
spacetime is expected to be stable in the complementary regime
$\mu\geq \sqrt{2}\cdot m\Omega$. This provides an {\it improved}
bound on the stability regime of the massive Klein-Gordon equation
in rotating Kerr spacetimes.

\bigskip
\noindent
{\bf ACKNOWLEDGMENTS}
\bigskip

This research is supported by the Carmel Science Foundation. I thank
Oded Hod, Yael Oren, Arbel M. Ongo and Ayelet B. Lata for helpful
discussions. I would also like to thank the anonymous referee for a
valuable comment regarding the Hilbert space considered in Refs.
\cite{Beyer1,Beyer2}.



\begin{thebibliography}{99}

\bibitem{RegWheel} T. Regge and J. A. Wheeler, Phys. Rev. {\bf 108}, 1063 (1957).

\bibitem{Vish} C.V. Vishveshwara, Nature {\bf 227}, 936 (1970).

\bibitem{PressTeu1} W. H. Press and S. A. Teukolsky, Astrophys. J. {\bf 185}, 649 (1973).

\bibitem{Teu} S. A. Teukolsky, Phys. Rev. Lett. {\bf 29}, 1114 (1972);
Astrophys. J. {\bf 185}, 635 (1973).

\bibitem{Hodp1} S. Hod, Phys. Rev. D {\bf 58}, 104022 (1998) [arXiv:gr-qc/9811032]; S. Hod,
Phys. Rev. D {\bf 61}, 024033 (2000) [arXiv:gr-qc/9902072]; S. Hod,
Phys. Rev. D {\bf 61}, 064018 (2000) [arXiv:gr-qc/9902073]; L.
Barack, Phys. Rev. D {\bf 61}, 024026 (2000); S. Hod, Phys. Rev.
Lett. {\bf 84}, 10 (2000) [arXiv:gr-qc/9907096]; R. J. Gleiser, R.
H. Price, and J. Pullin, Class. Quant. Grav. {\bf 25}, 072001
(2008); M. Tiglio, L. E. Kidder, and S. A. Teukolsky, Class. Quant.
Grav. {\bf 25}, 105022 (2008); S. Hod, Phys. Lett. B {\bf 666}, 483
(2008) [arXiv:0810.5419]; S. Hod, Phys. Rev. D 78, 084035 (2008)
[arXiv:0811.3806]; A. Zenginoglu and M. Tiglio, Phys. Rev.D {\bf
80}, 024044 (2009); A. J. Amsel, G.  T. Horowitz, D. Marolf, and M.
M. Roberts, J. High Energy Phys. 0909:044 (2009).

\bibitem{Mano} C. A. Manogue, Annals of Physics {\bf 181}, 261
(1988).

\bibitem{Grein} W. Greiner, B. M\"uller and J. Rafelski, {\it
Quantum electrodynamics of strong fields}, (Springer-Verlag, Berlin,
1985).

\bibitem{Zel} Ya. B. Zel`dovich, Pis`ma Zh. Eksp. Teor. Fiz. {\bf
14}, 270 (1971) [JETP Lett. {\bf 14}, 180 (1971)]; Zh. Eksp. Teor.
Fiz. {\bf 62}, 2076 (1972) [Sov. Phys. JETP {\bf 35}, 1085 (1972)].

\bibitem{Car} V. Cardoso, O. J. C. Dias, J. P. S. Lemos and S.
Yoshida, Phys. Rev. D {\bf 70}, 044039 (2004); Erratum-ibid. D {\bf
70}, 049903 (2004).

\bibitem{PressTeu2} W. H. Press and S. A. Teukolsky, Nature {\bf
238}, 211 (1972).

\bibitem{Dam} T. Damour, N. Deruelle and R. Ruffini, Lett. Nuovo
Cimento {\bf 15}, 257 (1976).

\bibitem{Zour} T. M. Zouros and D. M. Eardley, Annals of physics
{\bf 118}, 139 (1979).

\bibitem{Det} S. Detweiler, Phys. Rev. D {\bf 22}, 2323 (1980).

\bibitem{Furu} H. Furuhashi and Y. Nambu, Prog. Theor. Phys. {\bf 112}, 983
(2004).

\bibitem{Dolan} S. R. Dolan, Phys. Rev. D {\bf 76}, 084001 (2007).

\bibitem{CarYo}  V. Cardoso and S. Yoshida, JHEP 0507:009 (2005).

\bibitem{HodHod} S. Hod and O. Hod, Phys. Rev. D {\bf 81}, Rapid communication 061502
(2010) [arXiv:0910.0734].

\bibitem{Beyer1} H. R. Beyer, Commun. Math. Phys. {\bf 221}, 659
(2001).

\bibitem{Beyer2} H. R. Beyer, J. Math. Phys. {\bf 52}, 102502 (2011).

\bibitem{Kerr} R. P. Kerr, Phys. Rev. Lett. {\bf 11}, 237 (1963);
R. H. Boyer and R. W. Lindquist, J. Math. Phys. {\bf 8}, 265 (1967).

\bibitem{Heun} A. Ronveaux, {\it Heun's differential equations}.
(Oxford University Press, Oxford, UK, 1995).

\bibitem{Flam} C. Flammer, {\it Spheroidal Wave Functions} (Stanford
University Press, Stanford, 1957).

\bibitem{Fiz1} P. P. Fiziev, e-print arXiv:0902.1277.

\bibitem{Fiz2} R. S. Borissov and P. P. Fiziev, e-print
arXiv:0903.3617.

\bibitem{Fiz3} P. P. Fiziev, Phys. Rev. D {\bf 80}, 124001 (2009); P. P. Fiziev, Class. Quant. Grav. {\bf 27}, 135001
(2010).

\bibitem{Hartle} J. B. Hartle and D. C. Wilkins, Commun. Math. Phys. {\bf 38}, 47 (1974).

\bibitem{Abram} M. Abramowitz and I. A. Stegun, {\it Handbook of
Mathematical Functions} (Dover Publications, New York, 1970).

\bibitem{Teuk} S. A. Teukolsky, Phys. Rev. Lett. {\bf 29}, 1114 (1972);
Astrophys. J. {\bf 185}, 635 (1973).

\bibitem{Hodcen} S. Hod, Phys. Rev. Lett. {\bf 100}, 121101 (2008) [arXiv:0805.3873];
S. Hod, Phys. Rev. D {\bf 75}, 064013 (2007) [arXiv:gr-qc/0611004].

\bibitem{Stro} T. Hartman, W. Song, and A. Strominger, JHEP 1003:118 (2010).

\bibitem{Carter} B. Carter, Commun. Math. Phys. {\bf 10}, 280 (1968).

\bibitem{Notered} It should be noticed that the solutions of Eq.
(\ref{Eq9}) with the boundary conditions (\ref{Eq11})-(\ref{Eq12})
do not assume values in the natural Hilbert space that is associated
to the reduced Klein-Gordon equation of Refs. \cite{Beyer1,Beyer2}
due to their asymptotic behavior at $r_+$.

\bibitem{Will} L. E. Simone and C. M. Will, Class. Quantum Grav. {\bf 9}, 963 (1992).

\bibitem{ZhKo} R. A. Konoplya and A. Zhidenko, Phys. Rev. D {\bf 73}, 124040
(2006); Rev. Mod. Phys. {\bf 83}, 793 (2011).

\bibitem{Hodmassive} S. Hod, Phys. Rev. D {\bf 84}, 044046 (2011)
[arXiv:1109.4080].

\bibitem{Massfr}  Y. D\'ecanini, A. Folacci, and B. Raffaelli,
arXiv:1108.5076.

\bibitem{Noteunbound} It is worth noting that for unbound states (those with
$\omega>\mu$), one can deduce a stronger bound on the stability
region: $\mu\geq m\Omega$. This bound follows from the couple of
inequalities: (1) $\omega>\mu$ which characterizes the unbound
states, and (2) $\omega<m\Omega$ for the existence of superradiant
amplification. Thus, unbound states may be unstable only in the
restricted regime $\mu<m\Omega$.

\end{thebibliography}
\end{document}